\documentclass[12pt]{elsart}
\usepackage{amssymb}
\usepackage{epsfig}
\usepackage{fleqn}

\newcommand{\jref}[4]{{\it #1} {\bf #2} (#3) #4}
\newcommand{\eprint}[2]{{\tt #1}{#2}}

\newcommand{\MPLA}[3]{\jref{Mod.\ Phys.\ Lett.}{A#1}{#2}{#3}}

\newcommand{\NPB}[3]{\jref{Nucl.\ Phys.}{B#1}{#2}{#3}}

\newcommand{\PLB}[3]{\jref{Phys.\ Lett.}{B#1}{#2}{#3}}
\newcommand{\PR}[3]{\jref{Phys.\ Rep.}{#1}{#2}{#3}}
\newcommand{\PRD}[3]{\jref{Phys.\ Rev.}{D#1}{#2}{#3}}

\newcommand{\PRL}[3]{\jref{Phys.\ Rev.\ Lett.}{#1}{#2}{#3}}

\newcommand{\SJNP}[3]{\jref{Sov.\ J.\ Nucl.\ Phys.}{#1}{#2}{#3}}

\newcommand{\EPL}[3]{\jref{Europhys.\ Lett.}{#1}{#2}{#3}}
\newcommand{\JHEP}[3]{\jref{J.\ High\ Energy\ Phys. }{#1}{#2}{#3}}

\newcommand{\HEPTH}[1]{\eprint{hep-th/}{#1}}
\newcommand{\HEPPH}[1]{\eprint{hep-ph/}{#1}}

\begin{document}

\hoffset=0.5cm
\begin{frontmatter}
\begin{flushright}
HIP-2000-24/TH \\
hep-th/0005167
\end{flushright}

\vspace{2 cm}

\title{High-dimensional sources for the four-dimensional gravity}

\author{M. Chaichian$^{\rm a}$, M. Gogberashvili$^{\rm a,b}$ and
A.B. Kobakhidze$^{\rm a,b}$}
\address{$^{\rm a}$High Energy Physics Division, Department
of Physics, University of Helsinki and \\ Helsinki Institute of
Physics, P.O. Box 9, FIN-00014 Helsinki, Finland \\ $^{\rm
b}$Andronikashvili Institute of Physics, Georgian Academy of
Sciences \\ 6 Tamarashvili Str., 380077 Tbilisi, Georgia }

\begin{abstract}
We argue that a certain distribution of matter in higher
dimensions can provide the correct behaviour of gravity in four
dimensions. Some explicit examples illustrating the idea are
considered.
\end{abstract}
\end{frontmatter}


It is widely believed that a consistent unification of all
fundamental forces in nature would be possible within the
space-time with an extra dimensions beyond those four observed so
far. The absence of any signature of extra dimensions in current
experiments is usually explained by the compactness of extra
dimensions. Remarkably, the size of such dimensions can be
macroscopically large ($\sim 1$mm), if the only known particle
filling the extra dimensions is a graviton, while the other
particles are localized on the four-dimensional submanifold
(3-brane) embedded in higher-dimensional space-time \cite{1}.
Alternatively, one can consider high-dimensional space-time with
non-factorizable geometry where the four-dimensional Newton law is
correctly reproduced even in the case of non-compact extra
dimensions \cite{2,3} (for an earlier related works, see \cite{4}
and for some recent extensions and modifications, see
\cite{5,6,7,8,9,10}). The source for such a gravity is the 3-brane
embedded into the five-dimensional anti-de Sitter space-time with
the tension finely tuned to the value of the five-dimensional
negative cosmological constant \cite{2,3}.

In this paper we analyze more general high-dimensional sources
which indeed lead to the correct behaviour of gravity in four
dimensions. Let us start with five-dimensional Einstein's
equations:
\begin{equation}
\label{1}G_{AB}\equiv
R^{(5)}_{AB}-\frac{1}{2}g_{AB}R^{(5)}=k_{5}^{2}T_{AB},
\end{equation}
where $k_{5}^{2}=8\pi G_{N}^{(5)}=8\pi /M_{5}^{3}$ and
$A,B=0,...,4$. Taking the anzatz\footnote{Here we are interested
in the static solution. For non-static (cosmological) solutions
see the recent review \cite{11} and references therein.}
\begin{equation}
\label{2}
ds^2=a^2(x^{4})\eta_{\alpha\beta}dx^{\alpha}dx^{\beta}+(dx^{4})^{2},
\end{equation}
for the line-element of the five-dimensional space-time, which
respects the four-dimensional Poincar\'e invariance, Eqs.(\ref{1})
decompose as:
\begin{equation}
\label{3} a^{-2}G^{\alpha}_{\beta}+3\biggl [\frac{a''}{a}+\biggl
(\frac{a'}{a}\biggr )^{2}\biggr ]\delta^{\alpha}_{\beta}
=k_{5}^{2}T^{\alpha}_{\beta},
\end{equation}
\begin{equation}
\label{4} -a^{-2}R+6\biggl (\frac{a'}{a}\biggr
)^{2}=k_{5}^{2}T^{5}_{5},
\end{equation}
where $G^{\alpha}_{\beta}$ and $R$ are the  four-dimensional
Einstein tensor and Ricci scalar, respectively, built up from the
four-dimensional metric $\eta_{\alpha\beta}$ ($\alpha, \beta
=0,...,3$) and the prime denotes differentiation with respect to
the extra coordinate $x^{4}$. From (\ref{3}) and (\ref{4}) it is
obvious that in order to have correct Einstein's equations in four
dimensions (i.e. the correct four-dimensional gravity), we should
have
\begin{equation}\label{5}
G^{\alpha}_{\beta}=k_{4}^{2}\tau^{\alpha}_{\beta},
\end{equation}
where ordinary four-dimensional Newton's constant $k_{4}^{2}=8\pi
G_{N}=8\pi /M_{Pl}^{2}$ is related to the five-dimensional one
according to
\begin{equation}\label{6}
k_{5}^{2}=k_{4}^{2}\int a^{2}(x^{4})dx^{4}.
\end{equation}
Therefore, it is necessary to satisfy the following differential
equations:
\begin{equation}\label{7}
3\biggl [\frac{a''}{a}+\biggl (\frac{a'}{a}\biggr )^{2}\biggr
]\delta^{\alpha}_{\beta} =k_{5}^{2}\tilde{T}^{\alpha}_{\beta},
\end{equation}
\begin{equation}\label{8}
6\biggl (\frac{a'}{a}\biggr )^{2}=k_{5}^{2}\tilde{T}^{5}_{5}.
\end{equation}
Here we identify
$\tilde{T}^{\alpha}_{\beta}=T^{\alpha}_{\beta}-k^{2}_{4}k^{-2}_{5}a^{-2}
\tau^{\alpha}_{\beta}$ and
$\tilde{T}^{5}_{5}=T^{5}_{5}-k^{2}_{4}k^{-2}_{5}a^{-2}\tau^{\alpha}_{\alpha}$
with the energy-momentum stress tensor of the matter source in
extra space, while $\tau^{\alpha}_{\beta}$ is the energy-momentum
stress tensor of the four-dimensional matter. Usually, one chooses
a definite stress tensor that describes a certain distribution of
matter in extra dimensions and then looks for the solutions of the
system of Eqs. (\ref{7}) and (\ref{8}). Obviously, the non-trivial
solution of Eqs. (\ref{7}) and (\ref{8}) requires a certain
relation among the components of the chosen energy momentum tensor
$\tilde{T}^{A}_{B}$. For example, the original solution of Refs.
\cite{2,3}\footnote{The solution with negative exponent \cite{3}
corresponds to the case when the gravity is localized on the
3-brane at $x^{4}=0$. Such a 3-brane gravitationally repulses the
matter, thus an extra mechanism for the localization of matter on
the 3-brane is necessary to be imposed (for recent ideas of
localizing matter within the field-theoretic approach, see
\cite{12}). Alternatively, for the solution with a positive
exponent \cite{2} the 3-brane is gravitationally attractive, while
gravity itself is not localized on the 3-brane at $x^{4}=0$
(formally it is localized at infinity).},
\begin{equation}\label{9}
a^{2}=\exp(\pm 2\kappa|x^{4}|),
\end{equation}
arises for the energy-momentum stress tensor
\begin{equation}\label{10}
\tilde{T}^{\alpha}_{\beta}=-\biggl (\Lambda + \sigma \delta
(x^{4})\biggr )\delta^{\alpha}_{\beta},
\end{equation}
\begin{equation}\label{11}
 \tilde{T}^{5}_{5}=-\Lambda,
\end{equation}
if the five-dimensional cosmological constant $\Lambda$ ($\Lambda
< 0$) and the 3-brane\footnote{The 3-brane is considered
infinitely thin with a tension picked at $x^{4}=0$. For the
discussion of thick branes, see e.g. \cite{13}.} tension $\sigma$
are related as:
\begin{equation}\label{12}
 \sigma = \frac{\kappa}{k^{2}_{5}}~,~~~\kappa =
 k_{5}\sqrt{-\frac{\Lambda}{6}}.
\end{equation}

On the other hand, one can simply consider the Eqs. (\ref{7}) and
(\ref{8}) just as equations for the source energy-momentum tensor
$\tilde{T}^{A}_{B}$. Indeed, parameterizing $\tilde{T}^{A}_{B}$ as
\begin{equation}\label{13}
  \tilde{T}^{\alpha}_{\beta}=\frac{3}{k_{5}^{2}}F^{2}(x^{4})\delta^{\alpha}_{\beta},
\end{equation}
\begin{equation}\label{14}
  \tilde{T}^{5}_{5}=\frac{6}{k_{5}^{2}}Q^{2}(x^{4}),
\end{equation}
from (\ref{8}) one can easily obtain the solution
\begin{equation}\label{15}
  a=\exp \biggl \{ \int Q(x^{4})dx^{4}\biggr \}
\end{equation}
and from (\ref{7}), the equation related to the source functions
$F$ and $Q$:
\begin{equation}\label{16}
  Q'+2Q^{2}=F^{2}.
\end{equation}
The solution (\ref{15}) together with the relation (\ref{16}) is a
simple generalization of (\ref{9}) with (\ref{12}). Thus, we see,
that it is always possible to get a correct behaviour of gravity
in four dimensions by choosing the source functions (\ref{13}) and
(\ref{14}) satisfying (\ref{16}). However, in order to have a
fully consistent picture of all four-dimensional interactions one
should consider non-gravitational interactions as well in the
high-dimensional background determined by the source functions $F$
and $Q$. Indeed, each choice of the source functions $F$ and $Q$
requires further investigation in order to clarify in what extent
the physics could behave as being effectively four-dimensional,
not contradicting the current experiments. The usual way to
overcome this problem in the case of non-compact extra dimensions,
is to assume that the ordinary matter is localized in four
dimensions and thus does not fill the extra dimensions at all.
Here we assume that this is the case, not specifying the actual
mechanism responsible for such a localization. Now let us consider
some explicit solutions which essentially differ from those
already considered in the literature.
\begin{figure}[t]
\centerline{\protect\hbox{
\epsfig{file=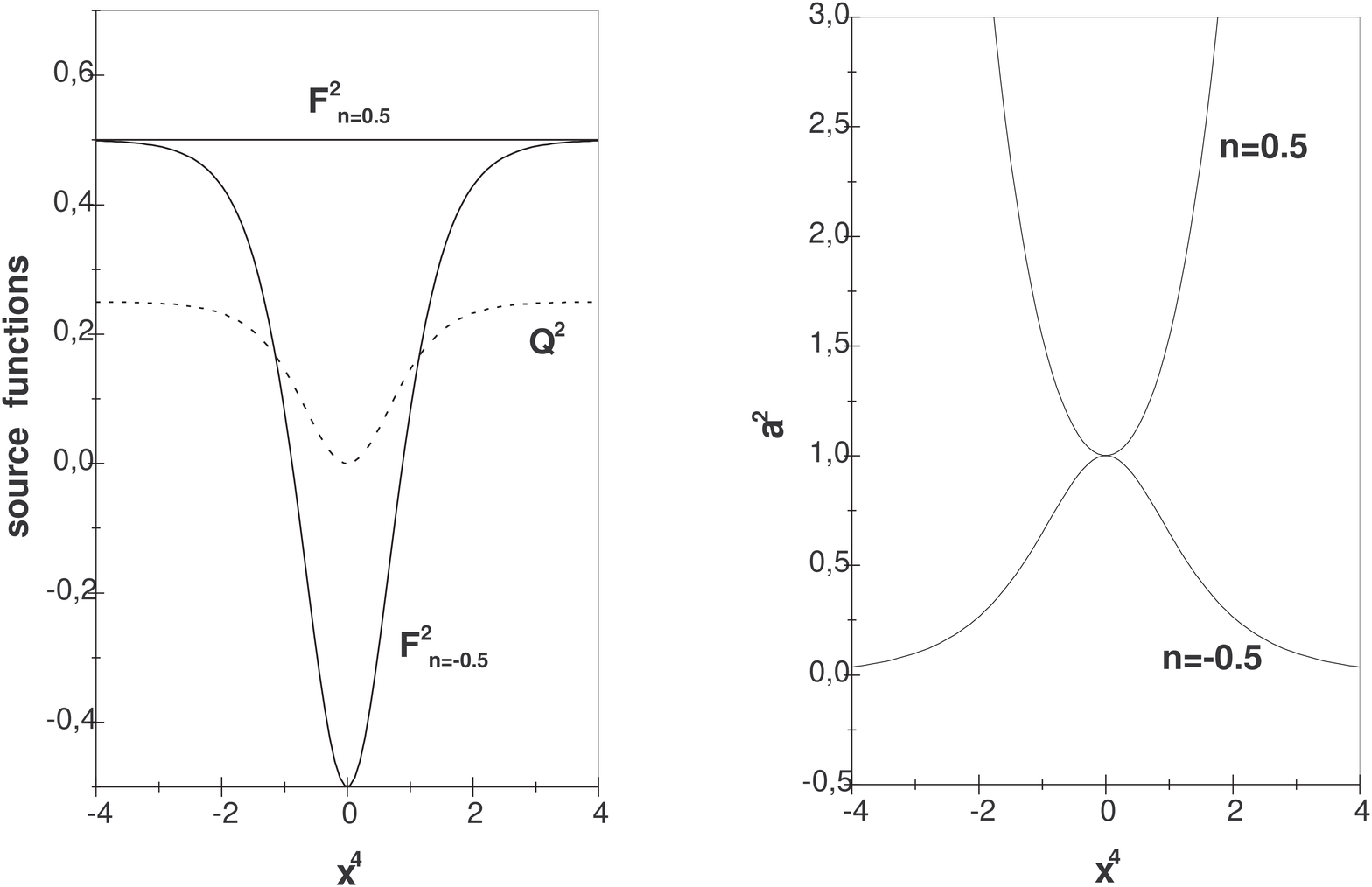,height=8.0cm,width= 14.0cm,angle=0}}}
\caption{\small Source functions (left) and conformal factors
(right) for the two particular cases $n=\pm 0.5$ of solutions A.
}
\label{fig1}
\end{figure}
\paragraph*{Solution A.} This solution corresponds to the choice
of the source functions as:
\begin{eqnarray}\label{17}
F^{2}=\frac{n}{\cosh^{2}(x^{4})}\biggl
[1+2n\sinh^{2}(x^{4})\biggr],  \nonumber \\
Q^{2}=n^{2}\tanh^{2}(x^{4}),
\end{eqnarray}
where $n$ is a constant parameter. The solution (up to the
integration constant which we set to be equal to 1) for the
conformal factor $a^2$ in (\ref{2}) is
\begin{equation}\label{18}
  a^{2}=\cosh^{2n}(x^{4}).
\end{equation}
For $n<0$ the solution (\ref{18}) is analogous to the solution of
Refs. \cite{3} with  graviton zero mode localized in four
dimensions at $x^{^4}=0$. For $n>0$ we obtain a solution which
grows as we go away from the point $x^{^4}=0$, thus providing an
attractive gravitational potential for trapping the matter in four
dimensions around $x^{^4}=0$ \cite{2}. We have plotted the
functions $F^{2}$, $Q^{2}$ and the conformal factor $a^{2}$ in
Fig. \ref{fig1} for $n=\pm 1/2$. Note, that for $n=+1/2$,
$F^{2}=1/2$ is constant, while $Q^{2}$ drops to its minimum at 
$x^{4}=0$. Such a distribution of matter is somewhat opposite to
the original scenarios of Refs. \cite{2,3} (see Eqs. (\ref{10})
and  (\ref{11})).
\begin{figure}[t]
\centerline{\protect\hbox{
\epsfig{file=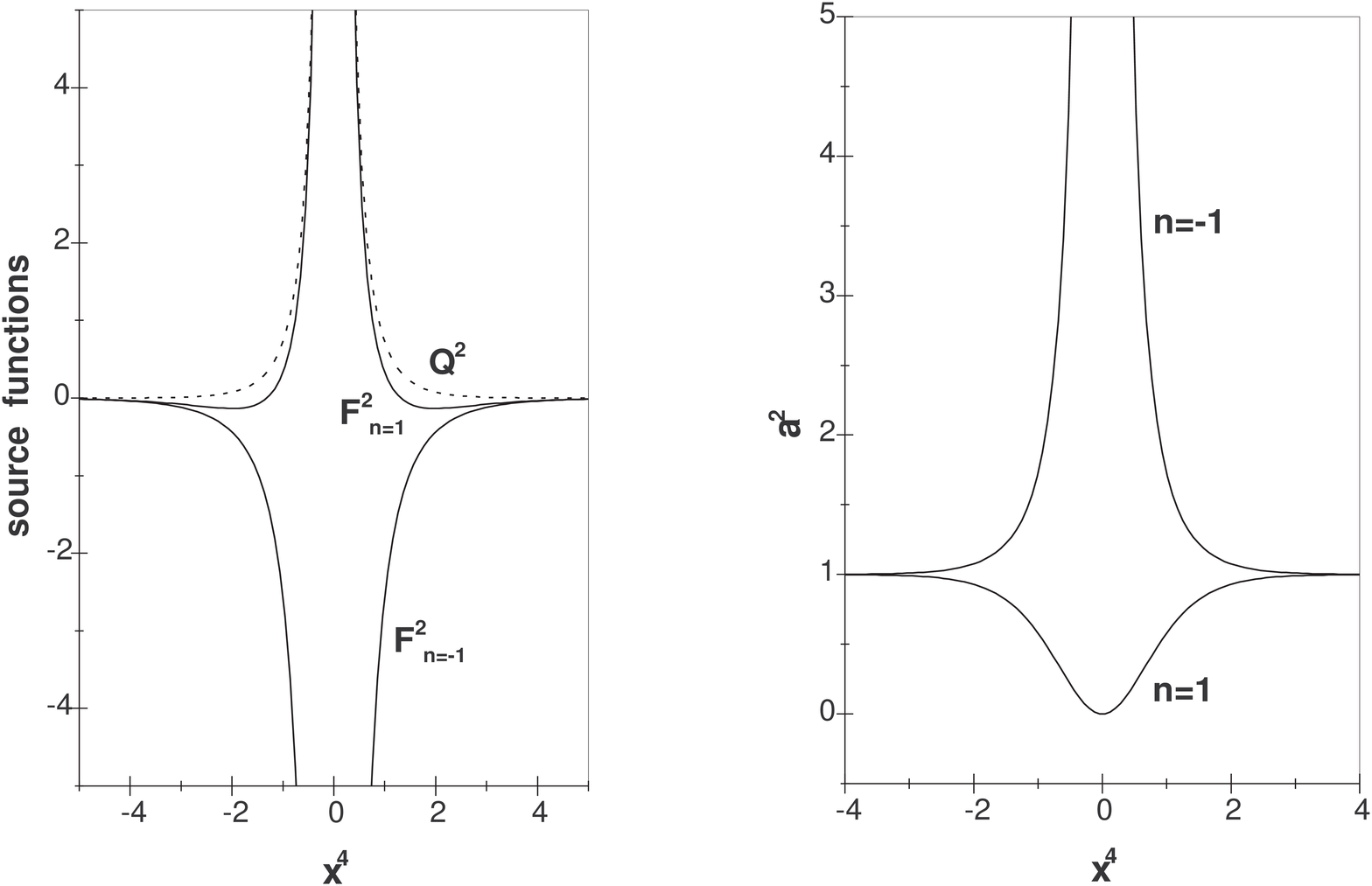,height=8.0cm,width= 14.0cm,angle=0}}}
\caption{\small The same as in Fig. \ref{fig1} for the two particular cases
$n=\pm 1$ of solutions B.
}
\label{fig2}
\end{figure}
\paragraph*{Solution B.} Now let us consider the following source
functions:
\begin{eqnarray}\label{19}
F^{2}=n^{2}{\rm csch}^{2}(x^{4})\biggl [2n-\cosh (x^{4})\biggr ],
\nonumber \\
Q^{2}=n^{2}{\rm csch}^{2}(x^{4}),
\end{eqnarray}
where $n$ is an arbitrary constant parameter again. The conformal
factor in this case is:
\begin{equation}\label{20}
  a^{2}=\tanh^{2n}(x^{4}).
\end{equation}
The source functions (\ref{19}) are in fact singular at $x^{4}=0$.
This leads to a singular behaviour of the  induced
four-dimensional metric ($a^{2}\eta_{\alpha \beta}$) at $x^{4}=0$
as well. For any real $n$, however, the conformal factor (\ref{20})
asymptotically approaches to one, while the source functions go to
zero (see Fig. \ref{fig2}). Thus far from the singular point
$x^{4}=0$, the space is essentially Minkowskian. One can imagine
that the ordinary four-dimensional matter is localized far from
the singularity, where the correct four-dimensional gravity could
be reproduced.
\begin{figure}[t]
\centerline{\protect\hbox{
\epsfig{file=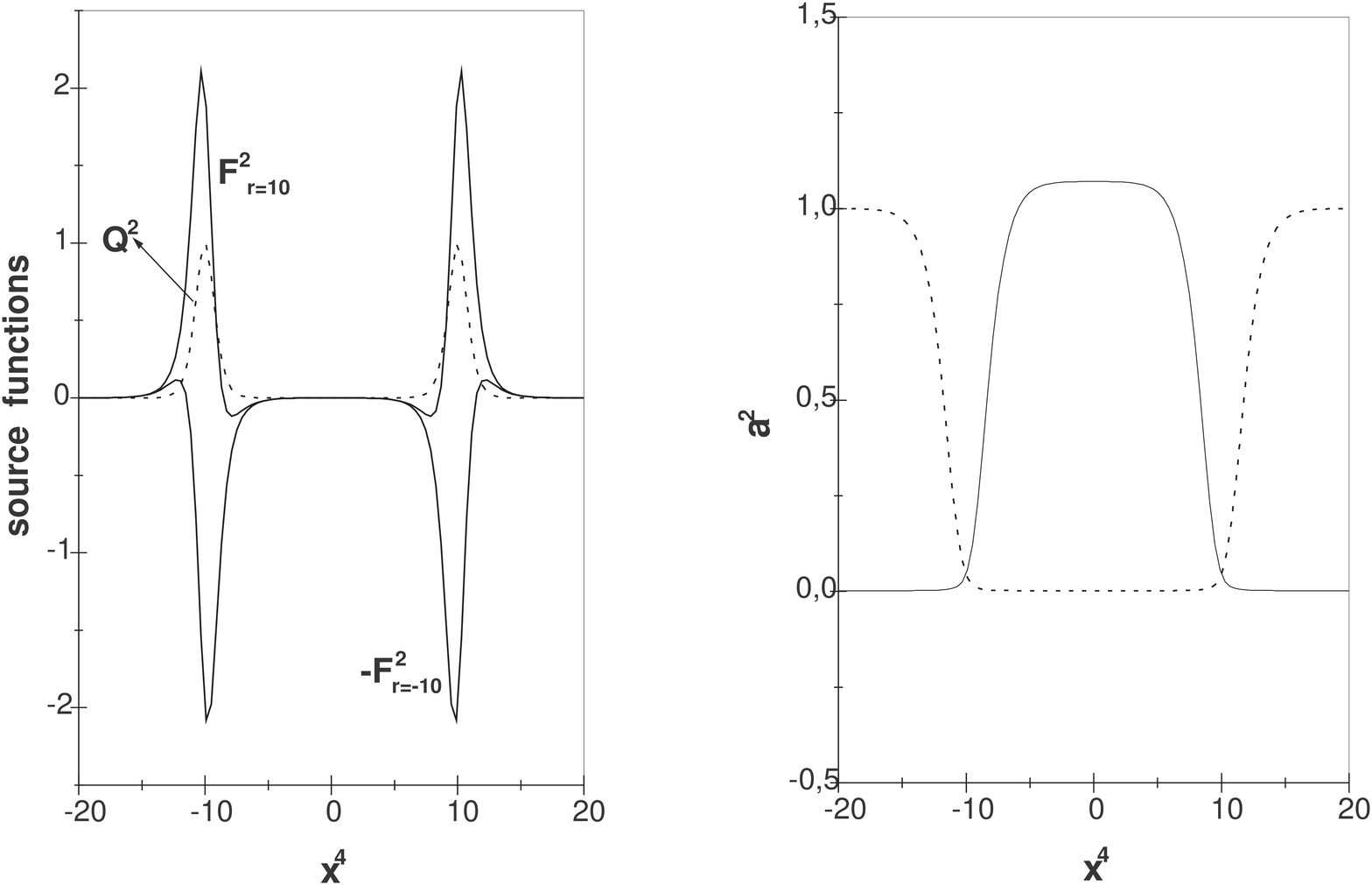,height=8.0cm,width= 14.0cm,angle=0}}}
\caption{\small The same as in Fig. \ref{fig1} for the two
particular cases $r=\pm 10$ (in an arbitrary unit of lenght) of
solutions C. Solid (dashed) curve on the right figure corresponds
to the positive (negative) $r$. } \label{fig3}
\end{figure}
\paragraph*{Solution C.}
Finally, let us consider some slightly more sophisticated sources:
\begin{eqnarray}\label{21}
  F^{2}=2\biggl [ {\rm sech}(x^{4}+r)-{\rm
 sech}(x^{4}-r)\biggr ]^2 \nonumber \\
 -{\rm sech}(x^{4}+r)\tanh(x^{4}+r)+{\rm
 sech}(x^{4}-r)\tanh(x^{4}-r), \nonumber \\
 Q^{2}=\biggl [ {\rm sech}(x^{4}+r)-{\rm sech}(x^{4}-r)\biggr ]^2,
\end{eqnarray}
where $r$ is a free parameter which determines the distance
between two peaks of the source functions $F^{2}$ and $Q^{2}$ (see
Fig. \ref{fig3}). We find that the conformal factor in this case
is given by
\begin{equation}\label{22}
  a^{2}=\exp \biggl \{4\arctan\biggl [ \tanh (\frac{x^{4}+r}{2})\biggr ]-4\arctan
  \biggl [\tanh (\frac{x^{4}-r}{2})\biggr ]\biggr \}.
\end{equation}

The asymptotic behaviours of the conformal factor and the source
functions in this case are similar to the previous ones, i.e. at
infinity we have the flat Minkowskian space. For positive $r$, in
the region $|x^{4}|<r$ the conformal factor is a decreasing
function as we go away from $x^{4}=0$ point. Therefore, like the
Refs. \cite{3} one can expect that gravity is localized at
$x^{4}=0$, while now one can place the matter on the "branes"
corresponding to the peaks (minima) of $F^{2}$ (see, Fig.
\ref{fig3}). This situation is similar to multi-brane models of
Refs. \cite{5}, so that one can also speculate on the solution to
the hierarchy problem. For negative $r$, the conformal factor
around $x^{4}=0$ behaves as in \cite{2}.

Clearly, plenty of other solutions with different source functions
$F$ and $Q$ could be considered as well. Here we have considered
relatively simple source functions with shapes which mimic
$\delta$-function type sources (thin branes). Obviously, in order
to make a final conclusion about the phenomenological implications
and the validity of such types of models, a more detailed analysis
similar to those \cite{14} done for the previously proposed models
\cite{2,3} has to be performed.

We conclude with the following comments. It is remarkable that,
while we have concentrated here on five-dimensional case, our
approach can be straightforwardly extended to any number of extra
dimensions. This is not the case for the proposals \cite{2,3}
which are valid for five dimensions and an extension to higher
dimensions is possible only within the framework of intersecting
branes \cite{6}, or within the sting-like objects in the case of
two extra dimensions \cite{7}. The next comment concerns the
source functions. Needless to say, that the source functions
considered above are chosen {\it ad hoc}, rather than obtained
from an underlying dynamics. In its turn it is an interesting and 
rather non-trivial task to obtain the desired distribution of
matter in higher dimensions from some more general theory that
ensures the correct behaviour of gravity in four dimensions. Here
we can only hope that such a theory can be constructed.


\paragraph*{Acknowledgement.} This work was supported by the Academy of Finland under the
Project No. 163394.


\end{document}